# Magnetic-Field-Induced Spin Nematicity in FeSe$_{1-x}$S$_x$ and FeSe$_{1-y}$Te$_y$ Superconductor Systems

Shaobo Liu[1,2], Jie Yuan[1,3,4], Sheng Ma[1,2], Zouyouwei Lu[1,2], Yuhang Zhang[1,2], Mingwei Ma[1,4], Hua Zhang[1,2], Kui Jin[1,2,3,4], Li Yu[1,2,4], Fang Zhou[1,2,4*], Xiaoli Dong[1,2,3,4*], and Zhongxian Zhao[1,2,3,4]

[1] *Beijing National Laboratory for Condensed Matter Physics and Institute of Physics, Chinese Academy of Sciences, Beijing 100190, China*
[2] *School of Physical Sciences, University of Chinese Academy of Sciences, Beijing 100049, China*
[3] *Key Laboratory for Vacuum Physics, University of Chinese Academy of Sciences, Beijing 100049, China*
[4] *Songshan Lake Materials Laboratory, Dongguan, Guangdong 523808, China*

\* Correspondence to: fzhou@iphy.ac.cn (F.Z.); dong@iphy.ac.cn (X.D.)

**Abstract**

The angular-dependent magnetoresistance (AMR) of the *ab* plane is measured on the single crystals of iron-chalcogenides FeSe$_{1-x}$S$_x$ ($x$ = 0, 0.07, 0.13 and 1) and FeSe$_{1-y}$Te$_y$ ($y$ = 0.06, 0.61 and 1) at various temperatures under fields up to 9 T. A pronounced twofold-anisotropic carrier-scattering effect is identified by AMR, and attributed to a magnetic-field-induced spin nematicity that emerges from the tetragonal normal-state regime below a characteristic temperature $T_{sn}$. This magnetically polarized spin nematicity is found to be ubiquitous in the isoelectronic FeSe$_{1-x}$S$_x$ and FeSe$_{1-y}$Te$_y$ systems, no matter whether the sample shows an electronic nematic order at $T_s \lesssim T_{sn}$, or an antiferromagnetic order at $T_N < T_{sn}$, or neither order. Importantly, we find that the induced spin nematicity shows a very different response to sulfur substitution from the spontaneous electronic nematicity: The spin-nematic $T_{sn}$ is not suppressed but even enhanced by the substitution, whereas the electronic-nematic $T_s$ is rapidly suppressed, in the FeSe$_{1-x}$S$_x$ sysytem. Furthermore, we find that the superconductivity is significantly suppressed with the enhancement of the induced spin nematicity in both FeSe$_{1-x}$S$_x$ and FeSe$_{1-y}$Te$_y$ samples.

The superconductivity in iron-based compounds emerges as the antiferromagnetic (AFM) order in their metallic parent compounds is suppressed,[1-4] analogous to the superconductivity in cuprates that is achieved by doping an AFM Mott insulator.[5] This supports the AFM fluctuations as the driving force for electron pairing. In iron chalcogenide systems of $FeSe_{1-x}S_x$-$FeSe_{1-y}Te_y$, the binary parent compound FeTe (with $y = 1$ and presence of interstitial Fe) shows a double-stripe AFM order with a ($\sim\pi/2,\sim\pi/2$) wave vector[6-8] at $T_N = 60 - 75$ K.[9-12] A dome-like superconducting regime appears upon suppression of the antiferromagnetism by substituting Se for Te.[2,9,10] At the other end of the systems, no long-range AFM order but the presence of single-stripe ($\pi$,0) AFM fluctuations[13] is observed in FeS ($x = 1$), which superconducts at $T_c \sim 5$ K.[14] The intermediate FeSe ($x = 0$) with $T_c \sim 9$ K is unique in that, while in absence of an AFM order, it displays an electronic nematic order associated with a tetragonal to orthorhombic structural transition upon cooling to $T_s \sim 90$ K.[15] This seems to raise a question whether the pairing is mediated still by the AFM fluctuations or by the nematic fluctuations. Nevertheless, previous inelastic neutron scatterings of FeSe have revealed in its tetragonal normal-state regime the isotropic ($\pi$,$\pi$) Néel AFM fluctuations that coexist with the anisotropic ($\pi$,0) ones,[16] and in its nematic regime the dominating stripe fluctuations.[16-19] This crossover from the significant isotropic to dominating anisotropic AFM fluctuations with cooling also manifests itself in the macroscopic properties. For instance, in the in-plane angular-dependent magnetoresistance (AMR)[20] and directional magnetic[21] measurements, a twofold anisotropy has been observed to simultaneously appear at the nematic transition of FeSe. Recently, nuclear magnetic resonance[22] and theoretical[23] studies of FeSe suggest a nematic state that not only involves the Fe $d_{xz}/d_{yz}$ orbitals but may also involve the $d_{xy}$ orbital, likely due to nontrivial spin–orbit coupling effect.[22,24] On the other hand, recent experimental studies of the isoelectronic $FeSe_{1-x}S_x$-$FeSe_{1-y}Te_y$ [25] and FeSe [26] systems suggest that the physics in the tetragonal environment, *i.e.* beyond the electronic nematic regime, is fundamental for the origin of superconductivity. Further investigating how the AFM fluctuations evolve and affect the superconductivity in the $FeSe_{1-x}S_x$ and $FeSe_{1-y}Te_y$ systems is highly desirable for a better understanding of the origin.

In this work, we systematically measure the angular-dependent magnetoresistance within the *ab* plane of the single crystal samples of $FeSe_{1-x}S_x$ ($x = 0, 0.07, 0.13, 1$) and $FeSe_{1-y}Te_y$ ($y = 0.06$, 0.61, 1). The in-plane AMR measurement has proved to be effective and efficient in probing the temperature-dependent anisotropy of the spin correlations present in the layered iron-based compounds.[20,27,28] As expected for the electronically nematic $FeSe_{1-x}S_x$ and magnetically double-stripe $Fe_{1.19}Te$ samples, a pronounced twofold anisotropy in their AMR is observed below a characteristic temperature $T_{sn}$. In the stoichiometric nematic FeSe, $T_{sn}$ coincides with $T_s$ of the electronic nematicity, while $T_{sn}$ is found to be well above $T_s$ in the substituted nematic $FeSe_{1-x}S_x$ as well as $T_N$ of the double-stripe antiferromagnetism in the parent $Fe_{1.19}Te$. The similar twofold anisotropy has been identified by AMR before in a bulk $Fe_{1-x}Se$ system and ascribed to the emergence of a magnetic-field-induced spin nematicity below $T_{sn}$.[20] Interestingly, however, such a twofold-symmetric AMR is also identified here in the non-magnetic tetragonal FeS and FeSe$_{1-y}$Te$_y$ ($y$ = 0.06, 0.61) samples. Importantly, we find that the isoelectronic substitution with sulfur does not suppress but even enhances $T_{sn}$ of the induced spin nematicity in the FeSe$_{1-x}$S$_x$ system, which contrasts sharply with its rapidly suppressed $T_s$ of the spontaneous electronic nematicity. Furthermore, we find that $T_c$ of the superconductivity is significantly suppressed with enhanced $T_{sn}$ of the induced spin nematicity in both FeSe$_{1-x}$S$_x$ and FeSe$_{1-y}$Te$_y$ samples. We discuss this suppression of superconductivity within the AFM-fluctuation-driven pairing scenario.

The in-plane angular-dependent magnetoresistance and zero-field resistivity were measured on a Quantum Design PPMS-9 system with a four-probe configuration. The AMR data at various temperatures under magnetic fields (up to 9 T) parallel to the *ab* plane were obtained by rotating the single crystal samples about their *c* axis. The x-ray $\phi$-scan experiments of the (103) plane were carried out at room temperature on a diffractometer (Rigaku SmartLab, 9 kW) equipped with two Ge (220) monochromators. Two series of the single crystal samples of FeSe$_{1-x}$S$_x$ ($x$ = 0, 0.07, 0.13, 1) and FeSe$_{1-y}$Te$_y$ ($y$ = 0.06, 0.61, 1) are studied here. For all the samples except for FeS, the crystal growth, phase identification, crystallographic characterization, chemical composition, and the electronic and superconducting properties have been reported elsewhere.[25] The corresponding characterizations of the FeS single crystal grown by a hydrothermal method[29] are presented in Fig. S1 of the supplementary material (SM). The temperature-dependent scaled resistivity and its first derivative of AFM parent Fe$_{1.19}$Te ($y$ = 1) with $T_N$ ~ 59 K are given in Fig. S2 of the SM. Its antiferromagnetic transition and concomitant structural transition[9,11,12] are manifest in the pronounced resistive anomaly at $T_N$ (Fig. S2).

The relative directions of the in-plane magnetic field and electric current with respect to the tetragonal *a* or *b* axis (the diagonal directions of the Fe square lattice) in our AMR measurements were determined by the room-temperature x-ray $\phi$ scans of the (103) plane. The experimental uncertainty is usually about 5°. As seen from Figs. 1(a-c), the $\phi$-scan data of three representative samples (FeS, FeSe, FeSe$_{0.39}$Te$_{0.61}$) exhibit four successive peaks with an equal interval of 90°, consistent with the $C_4$ lattice rotational symmetry. The similar $\phi$-scan results are obtained in other FeSe$_{1-x}$S$_x$ ($x$ = 0.07) and FeSe$_{1-y}$Te$_y$ ($y$ = 0.06, 1) samples and given in Fig. S3. Here we define $\alpha$ as the angle between the current *I* and *a* (or *b*) axis, and $\theta$ as the angle between the field *H* and *b* (or *a*) axis, as schematically illustrated in Figs. 1(g) and 1(h) for two of the different current directions, $\alpha$ = 0° and 45°, respectively. At current angle $\alpha$ = 45°, *I* is along the nearest-neighbour (NN) Fe-Fe direction, with field angle $\theta$ = 45° (or 225° *etc.*) corresponding to $H//I$. At $\alpha$ = 0°, *I* is along the next NN (NNN) Fe-Fe direction, with $\theta$ = 0° (or 180° *etc.*) corresponding to $H \perp I$.

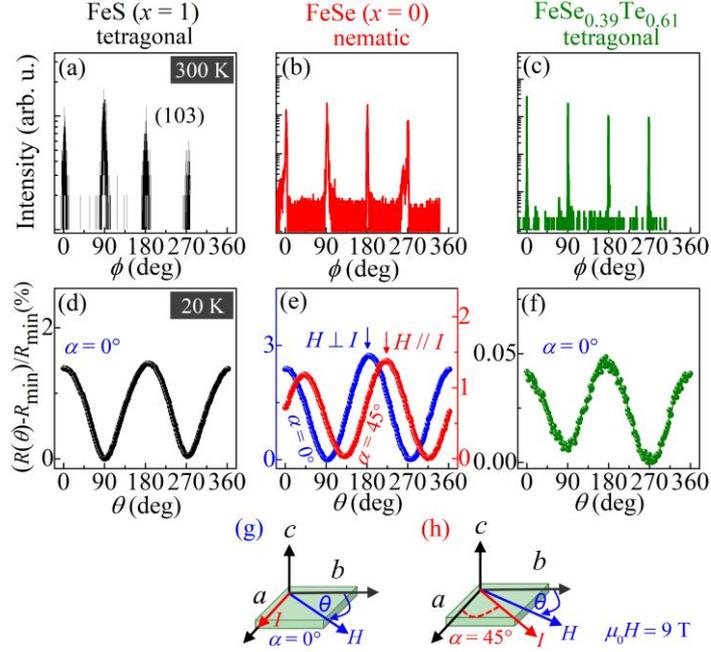

Fig. 1. The results of room-temperature x-ray $\phi$ scans of the (103) plane for (a) FeS, (b) FeSe, and (c) FeSe$_{0.39}$Te$_{0.61}$ agree with the $C_4$ lattice rotational symmetry and are used to calibrate the relative orientations of AMR measurements. The data of anisotropic AMR measured at temperature of 20 K and magnetic field of 9 T are presented here as the ratios of $\Delta R(\theta)/R_{\min}$ for (d) tetragonal FeS, (e) nematic FeSe ($T_s \sim 89$ K), and (f) tetragonal FeSe$_{0.39}$Te$_{0.61}$. Two of the different configurations of AMR measurement, with the current angle $\alpha = 0°$ (g) and $45°$ (h), are schematically illustrated.

In Figs. 1(d-f), the AMR results at the low temperature $T = 20$ K and the magnetic field $\mu_0 H = 9$ T are presented as the ratios of $\Delta R(\theta)/R_{\min} = [R(\theta)-R_{\min}]/R_{\min} \times 100\%$ for the three samples. Here $R_{\min}$ is the minimum resistance when the magnetic field rotates from $\theta = 0°$ to $360°$. The twofold anisotropy in AMR is evident. This $C_2$ rotational symmetry in all our samples is noticeable from the polar plots of their $\Delta R(\theta,T)/R_{\min}(T)$ shown in Fig. 2, with Figs. 2(a) and 2(b-d) being the data for tetragonal ($x = 1$) and nematic ($x = 0.13, 0.07, 0$) FeSe$_{1-x}$S$_x$ samples, respectively, Figs. 2(e) ($y = 0.06$) and 2(f) ($y = 0.61$) for tetragonal FeSe$_{1-y}$Te$_y$, and Fig. 2(g) for AFM parent Fe$_{1.19}$Te ($y = 1$). We emphasize that the twofold-symmetric AMR emerges no matter whether the system is nematically ordered (*e.g.*, FeSe with $C_4$ to $C_2$ symmetry breaking at $T_s$), or magnetically ordered (Fe$_{1.19}$Te with $C_4$ symmetry broken at $T_N$), or show neither of the orders (*e.g.*, FeS and FeSe$_{0.39}$Te$_{0.61}$ with persistent $C_4$ lattice symmetry). The presence or absence of the electronic nematicity in the samples has been characterized in our recent study.[25]

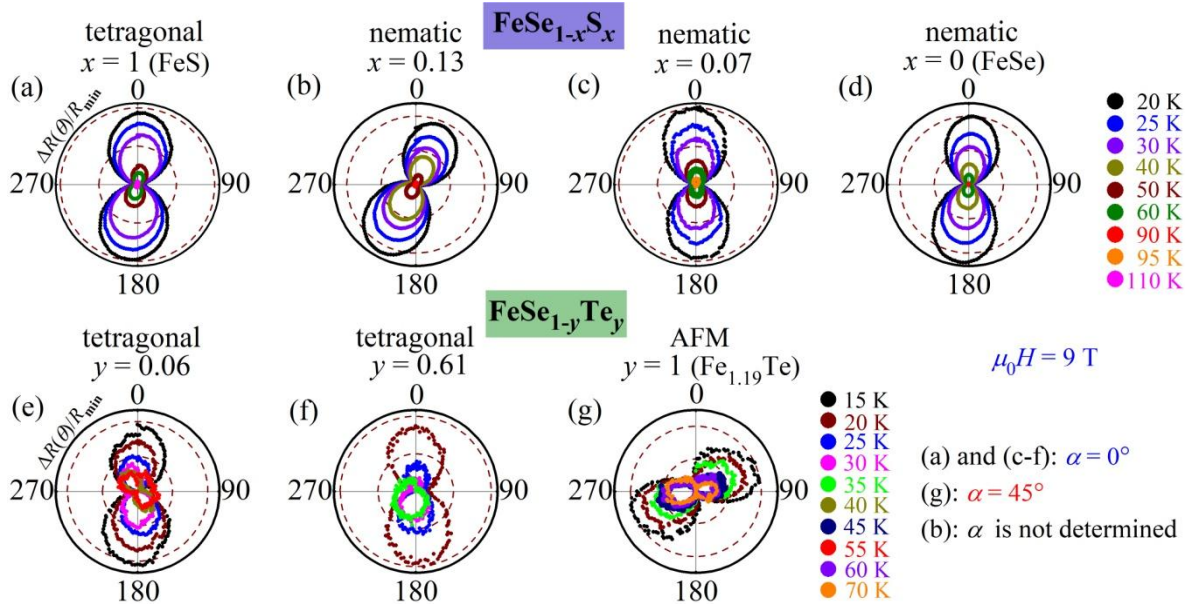

Fig. 2. The polar plots of $\Delta R(\theta,T)/R_{min}(T)$ measured under magnetic field of 9 T for (a-d) FeSe$_{1-x}$S$_x$ ($x$ = 1, 0.13, 0.07, 0) and (e-g) FeSe$_{1-y}$Te$_y$ ($y$ = 0.06, 0.61, 1) samples, respectively. The current angle is $\alpha = 0°$ for (a) and (c–f), and $\alpha = 45°$ for (g). For $x = 0.13$ in (b), the relative orientation of AMR measurement was not checked by x-ray $\phi$ scan. For AFM Fe$_{1.19}$Te in (g), the field angle $\theta$ corresponding to the maximum AMR direction twists at $T \gtrsim 45$ K, and the size of the twist angle is increased to ~40° at $T = T_{sn} \sim 70$ K. See main text for discussion. In (a) and (d), the field angle $\theta \sim 5°$ corresponds to $H \perp I$. This angle off-set from $\theta = 0°$ is due to the experimental uncertainty in setting up the rotating rod.

In order to quantify the temperature-dependent crossover from the isotropic to anisotropic carrier scatterings, we performed the AMR measurements at various temperatures. In Figs. 3(a-d) and 3(e-g), we present the temperature dependences of the maximum $\Delta R(T)/R_{min}(T)$ of AMR at the field of 9 T for the two series of samples of FeSe$_{1-x}$S$_x$ ($x$ = 0, 0.07, 0.13, 1) and FeSe$_{1-y}$Te$_y$ ($y$ = 0.06, 0.61, 1), respectively. It is obvious that the twofold anisotropy in AMR develops below a well-defined characteristic temperature $T_{sn}$, and this anisotropy becomes strongly enhanced as the temperature is further lowered. In FeSe, it happens that $T_{sn}$ (~90 K) coincides with $T_s$ (~89 K) of its electronic nematicity, exactly the same as the previous AMR study of the Fe$_{1-x}$Se system.[20] In Fe$_{1.19}$Te, however, the anisotropic AMR is found to set in at a $T_{sn} \sim 70$ K, which is well above $T_N \sim 59$ K of its double-stripe antiferromagnetism. Interestingly, in the non-magnetic tetragonal FeS with the lowest $T_c = 4.8$ K [Fig. S1(c)], the twofold-symmetric AMR is identified below a $T_{sn}$ as high as 110 K. This $T_{sn} \sim 110$ K is far above the measuring temperature (4 K) of the neutron scatterings revealing the single-stripe AFM fluctuations.[13] In the non-magnetic tetragonal FeSe$_{1-y}$Te$_y$, the twofold symmetry is observed below $T_{sn} \sim 40$ K for $y = 0.06$ ($T_c = 6.7$ K) and $T_{sn} \sim 30$ K for $y = 0.61$ ($T_c = 14$ K). We note that, although the AMR anisotropy of FeSe$_{1-y}$Te$_y$ is similarly pronounced to that of FeSe$_{1-x}$S$_x$, differences exist between them. The $T_{sn}$ ($\lesssim 70$ K) of FeSe$_{1-y}$Te$_y$ samples is found to be much lower than the $T_{sn}$ ($\gtrsim 90$ K) of FeSe$_{1-x}$S$_x$ samples. Moreover, the FeSe$_{1-y}$Te$_y$ samples [Figs. 3(e-g)] exhibit the anisotropic AMR signals about one to two orders of magnitude weaker than those of the FeSe$_{1-x}$S$_x$ samples [Figs. 3(a–d)], despite having their much more sharp isotropic-to-anisotropic crossover below $T_{sn}$ than the FeSe$_{1-x}$S$_x$ samples. These phenomena could be related to the facts that FeSe$_{1-y}$Te$_y$ is derived from parent Fe$_{1.19}$Te with the complex double-stripe antiferromagnetism and magnetic frustrations,[8] and both isotropic and anisotropic AFM fluctuations are present in substituted FeSe$_{1-y}$Te$_y$ (at, *e.g.*, $y = 0.62$), as revealed by the neutron scatterings.[2] In Figs. 3(a–g), the temperature regions of the twofold-anisotropic AMR, electronic nematicity, and double-stripe antiferromagnetism are marked by the pink, purplish-blue, and green shadings, respectively. Particularly for FeSe in Fig. 3(d), the coexisting Néel and stripe AFM fluctuations in the tetragonal regime,[16] as well as the dominating stripe ones in the nematic regime,[16-19] as revealed by the previous neutron studies are also indicated. Additionally, three of the neutron-scattering temperatures for FeSe [16] and FeS [13] are indicated in Fig. 4.

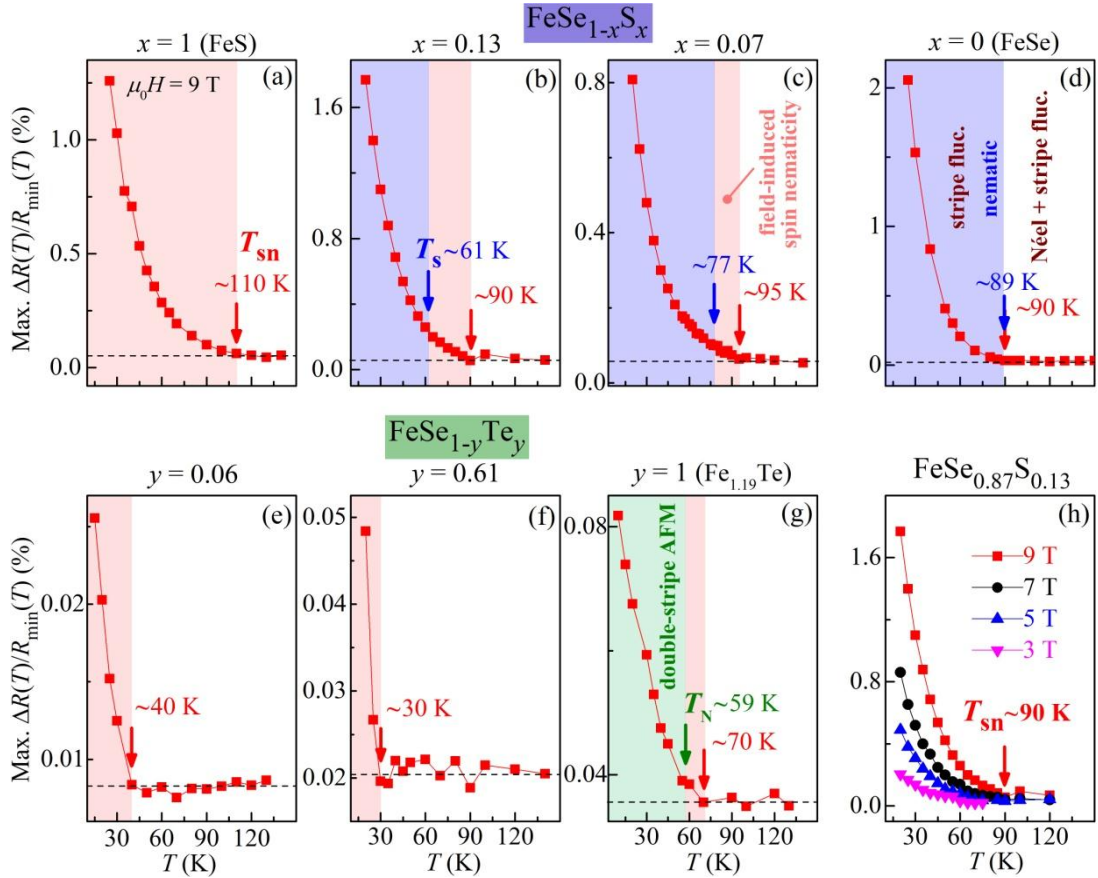

Fig. 3. (a–d) and (e–g) show the temperature dependences of the maximum $\Delta R(T)/R_{min}(T)$ of anisotropic AMR at the magnetic field of 9 T for FeSe$_{1-x}$S$_x$ ($x = 1, 0.13, 0.07, 0$) and FeSe$_{1-y}$Te$_y$ ($y = 0.06, 0.61, 1$) samples, respectively. The red, blue, and green arrows indicate the temperatures of $T_{sn}$, $T_s$, and $T_N$, respectively. For FeSe in (d), the coexisting Néel and stripe AFM fluctuations in the tetragonal regime,[16] as well as the dominating stripe ones in the nematic regime,[16-19] are also indicated. (h) The temperature-dependent maximum $\Delta R(T)/R_{min}(T)$ measured under various fields from 3 to 9 T for nematic FeSe$_{0.87}$S$_{0.13}$ ($x = 0.13$). Also see Fig. S5 for the extracted field dependences of the maximum $\Delta R(T)/R_{min}(T)$ at given $T < T_{sn}$.

The reproducibility of the twofold-symmetric AMR at $T < T_{sn}$ was also checked out at different current directions ($\alpha$'s) on FeSe. Fig. 1(e) shows the results obtained with the configurations of current $\alpha = 0°$ and $45°$. The results obtained on one and the same sample at $\alpha = 0°$, $22°$ and $90°$ are given in Fig. S4. The $C_2$ symmetry of AMR is well reproducible regardless of the different current directions. At the current $\alpha = 0°$ or $90°$ (the NNN Fe-Fe directions), the maximum and minimum of $\Delta R(\theta,T)/R_{min}(T)$ are reached with $H \perp I$ and $H // I$, respectively, when $H$ is parallel to the orthogonal NNN Fe-Fe directions [see Figs. 1(e), 1(g) and S4(c), S4(f) for $\alpha = 0°$ with the corresponding field $\theta = 180°$ (maximum) and $90°$ (minimum), and Figs. S4(b), S4(e) for $\alpha = 90°$]. Additionally, at the current $\alpha = 22°$ closer to the NNN Fe-Fe direction, as shown in Figs. S4(a) and S4(d), $\Delta R(\theta,T)/R_{min}(T)$ is maximized and minimized again when $H$ is parallel to the orthogonal NNN Fe-Fe directions (with the $\theta = 180°$ and $90°$, respectively), but the angles between $H$ and $I$ are changed to $<H, I> = 90° - 22°$ and $22°$, respectively (note that $<H, I> = 90°$ and $0°$ at $\alpha = 0°$). Interestingly, a discontinuity is observed upon the current $\alpha = 45°$, i.e. with $I$ along the NN Fe-Fe direction. In this case [Figs. 1(e) and 1(h)], the maximum and minimum of $\Delta R(\theta,T)/R_{min}(T)$ are observed at the reversed $H // I$ and $H \perp I$ (as compared to the case of $\alpha = 0°$), respectively, but not at $<H, I> = 90° - 45°$ and $45°$ (as extrapolated from the cases of $\alpha = 0°$ and $22°$), when $H$ is parallel to the orthogonal NN Fe-Fe directions [with the corresponding field $\theta = 225°$ (or $45°$) and $135°$ (i.e., $225° - 90°$), respectively]. Here we first discuss the implication of the maximum/minimum carrier-scattering effects; the abrupt change in the magnetic polarization effect upon the current-direction change to $\alpha = 45°$ from $\alpha = 0°$ will be discussed later. In Fig. 1(e), the maximum and minimum scattering effects observed at $H \perp I$ and $H // I$, respectively, for $\alpha = 0°$ imply that the spin arrays are aligned by the magnetic field with their directions of the inherent antiferromagnetic/ferromagnetic (FM) correlations between neighboring spins being mainly perpendicular/parallel to $H$. For $\alpha = 45°$, in contrast, the spin arrays are aligned by the field with their directions of the AFM/FM correlations between neighboring spins being mainly parallel/perpendicular to $H$, giving rise to the maximum and minimum scatterings at $H // I$ and $H \perp I$, respectively.

For the double-stripe AFM $Fe_{1.19}Te$ shown in Fig. 2(g), the maximum of its $\Delta R(\theta,T)/R_{min}(T)$ measured at current $\alpha = 45°$ is observed at the field $\theta$ (~$45°$) almost in the NN Fe-Fe direction, i.e. with $H$ almost parallel to $I$, at lower temperatures $\lesssim 45$ K. At higher temperatures $\gtrsim 45$ K, however, the maximum direction substantially deviates from the NN Fe-Fe direction by an angle up to ~$40°$ (at $T = T_{sn} \sim 70$ K). Note that the temperature $T \sim 45$ K is well *below* $T_N \sim 59$ K. Such a large twist angle cannot be ascribed to the experimental uncertainty. This may be explained by the combined effects of the spin interactions (that are frustrated[8]) and the magnetic polarization in $Fe_{1.19}Te$. When the spin correlation is strong, the magnetic field could only partially polarize the dominating anisotropic spin-correlated state. The polarization effect would change with temperature due to thermal fluctuations.

In our earlier AMR study of the $Fe_{1-x}Se$ system,[20] the similar twofold-anisotropic scattering effect was attributed to a spin nematicity formed below $T_{sn}$ due to the magnetic polarization effect, and an Ising-like order parameter was argued for the spin nematicity (see ref. 20 and references therein). In Fig. 3(h), the temperature dependences of maximum $\Delta R(T)/R_{min}(T)$ measured at various fields ≤ 9 T are plotted together for $x = 0.13$ $FeSe_{0.87}S_{0.13}$, which shows the electronic-nematic $T_s$ (~61 K) much lower than the spin-nematic $T_{sn}$ (~90 K). The magnetically induced nature of the spin nematicity is evident from Fig. 3(h). As the magnetic field strength is reduced, the twofold anisotropy in AMR at given $T < T_{sn}$ is correspondingly weakened. At the lower magnetic field of 3 T, the spin nematicity is just discernible by AMR. We note here that the appearance of the field-polarized spin nematicity below $T_{sn}$ in FeSe corresponds to the presence of dominating stripe AFM fluctuations at lower $T < T_s$ (~$T_{sn}$) as have been revealed by the neutron scatterings.[16-19] The disappearance of the field-polarized spin nematicity above $T_{sn}$ in turn corresponds to the presence of significant isotropic AFM fluctuations at higher $T > T_s$.[16] The anisotropic spin fluctuations are also present in FeS [13] and $FeSe_{1-y}Te_y$ [2,3] at low temperatures. The quickly increasing maximum $\Delta R(T)/R_{min}(T)$ with cooling as observed at $T < T_{sn}$ (Fig. 3) is consistent with the presence of the anisotropic spin correlations at low $T$. Moreover, the common twofold symmetry of AMR shared by all the $FeSe_{1-x}S_x$ and $FeSe_{1-y}Te_y$ samples suggests the common underlying physics.

Interestingly, as described and discussed above, there also appear the unusual phenomena in AMR of FeSe. While the same characteristic feature of the magnetic polarizations is observed with current $I$ along the orthogonal NNN Fe-Fe directions ($\alpha = 0°$ or $90°$), the polarization effects of field $H$ on the AFM/FM spin arrays are found to be abruptly changed upon $I$ along the NN Fe-Fe direction ($\alpha = 45°$). The maximum scatterings at current $\alpha = 0°$ and $90°$ can be interpreted to be due to the field-aligned AFM spin arrays mainly perpendicular to $H$ ($\perp I$), whereas those at $\alpha = 45°$ due to the field-aligned AFM spin arrays mainly parallel to $H$ (//$I$). This abrupt change upon $\alpha = 45°$ appears to be difficult to comprehend if the induced spin nematicity is contributed purely and simply by Fe $d_{xz}/d_{yz}$ orbitals, within the usual nematic picture. Therefore, such distinctly different magnetic polarization effects on the spin arrays observed at the currents along the nearest-neighbour ($\alpha = 45°$) and next nearest-neighbour ($\alpha = 0°/90°$) Fe-Fe directions deserve further investigation by taking account of the orbital-selective effect. For instance, the in-plane-lying $d_{xy}$ orbital may also be involved in the induced spin nematicity. Recent experiment[22] and theory[23] also suggest the possibility of $d_{xy}$ contribution to nematicity.

Further, we recall that, in magnetically ordered iron pnictides $SrFe_2As_2$ [27] and $BaFe_{2-x}Co_xAs_2$ [28] with $T_N \sim 200$ K and 138 K ($BaFe_2As_2$ at $x = 0$), respectively, the similar $C_2$-symmetric AMR was also observed and attributed to the occurrence of single-stripe antiferromagnetism commonly present in iron pnictides.[30] Once the stripe AFM order has disappeared (e.g., at $x > 0.2$ in $BaFe_{2-x}Co_xAs_2$, see ref. [28]), the twofold anisotropy in AMR simultaneously vanishes as well. On the other hand, previous neutron studies have detected the presence of local magnetic moments in FeS,[13] FeSe,[16] and FeSe$_{0.65}$Te$_{0.35}$[31] systems without magnetic order, in addition to the itinerant contribution. Moreover, an $S = 1$ paramagnet of FeSe,[16] and nontrivial spin–orbit coupling effect in FeSe [22,24] and FeSe$_{0.5}$Te$_{0.5}$,[32] have been suggested. Here, in the non-magnetic tetragonal FeS, the field-induced spin nematicity is significant at the high $T_{sn}$ ~ 110 K so that it is well detectable by AMR. This is possibly because the crystal lattice of stoichiometric FeS is most contracted [see Fig. S1(b) and ref. 25], thus its interlayer coupling should be strongest among the FeSe$_{1-x}$S$_x$ and FeSe$_{1-y}$Te$_y$ systems. Although a long-range AFM order fails to establish in the FeSe$_{1-x}$S$_x$ samples showing higher $T_{sn}$ (≳90 K) due to strongly frustrated magnetism,[8,16,33-39] it indeed occurs in the strongly correlated parent Fe$_{1.19}$Te at $T_N$. However, $T_{sn}$ (~70 K) of its induced spin nematicity is found to be well above $T_N$ (~59 K) of its spontaneous double-stripe antiferromagnetism. This means that incipient anisotropic spin correlations have already well developed in Fe$_{1.19}$Te at higher $T = T_{sn}$ (>$T_N$) in the presence of spin–orbit coupling effect, as a precursor of the double-stripe order eventually occurring at lower $T = T_N$. Overall, the magnetic-field-induced spin nematicity below $T_{sn}$ as reflected in the AMR measurement is ubiquitous in the FeSe$_{1-x}$S$_x$ and FeSe$_{1-y}$Te$_y$ systems, no matter whether the sample is nematically/magnetically ordered or not (Fig. 3). Therefore, it turns out that the enhancement of the anisotropy in AFM correlations below $T_{sn}$ is an inherent property of the systems.

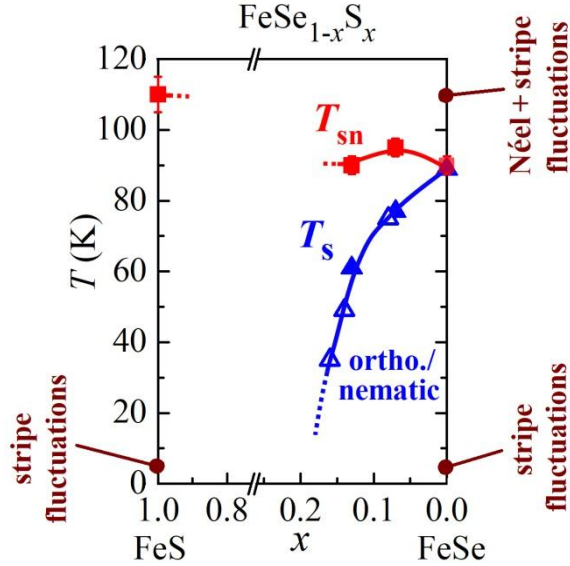

Fig. 4. Local $T$ vs $x$ phase diagram of the FeSe$_{1-x}$S$_x$ system. Here the data from ref. [40] are also included and denoted by the hollow blue triangles. Three of the temperatures of previous neutron experiments of FeSe [16] and FeS [13] are marked.

Having identified the ubiquitous presence of the magnetically induced spin nematicity, now we switch to further discussion of the implication for superconductivity. First of all, as can be seen from the local phase diagram of FeSe$_{1-x}$S$_x$ system plotted in Fig. 4, the isoelectronic substitution with sulfur strongly suppresses the electronic nematicity in FeSe$_{1-x}$S$_x$ samples. Their $T_s$ is reduced from ~89 K ($x = 0$; $T_c$ = 8.5 K), ~77 K ($x = 0.07$; $T_c$ = 10.3 K), to ~61 K ($x = 0.13$; $T_c$ = 10.8 K).[25] In contrast, their spin-nematic $T_{sn}$ is found to change little with substitution $x$ (there is only a small increase of $T_{sn}$ by ~5 K at $x = 0.07$). As a result, $T_{sn}$ of the induced spin nematicity is located well above $T_s$ of the spontaneous electronic nematicity in the substituted FeSe$_{1-x}$S$_x$ ($0.13 \geq x > 0$) samples. Particularly in the left-end stoichiometric FeS with the maximized substitution ($x = 1$), the spin-nematic $T_{sn}$ is even enhanced to ~110 K against the completely suppressed electronic nematicity. These experimental observations demonstrate that the isoelectronic sulfur substitution significantly reduces the electron-electron correlation due to the lattice contraction with increasing $x$ [25,41,42] [see Fig. S1(b) for $x = 1$], hence it directly suppresses the spontaneous electronic nematicity, rather than the induced spin nematicity. In this sense, it is very likely that the spin-correlation channel is separate from the electron-correlation channel.

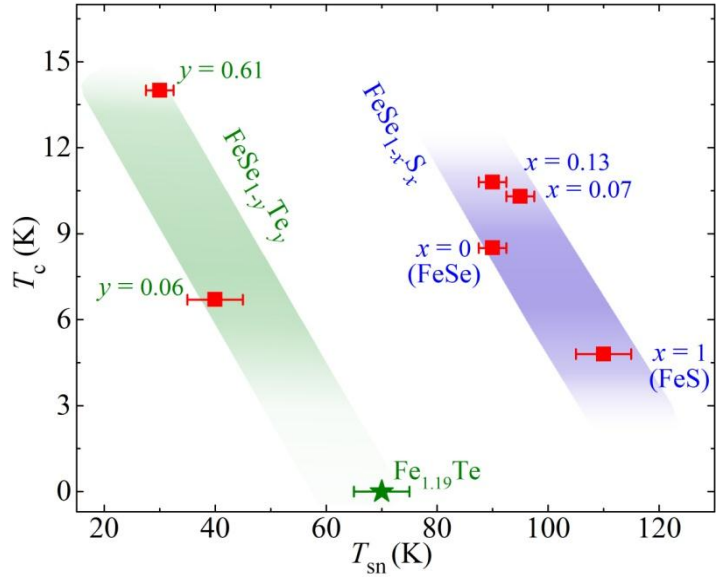

Fig. 5. $T_c$ of the superconductivity vs $T_{sn}$ of the field-induced spin nematicity for isoelectronic FeSe$_{1-x}$S$_x$ and FeSe$_{1-y}$Te$_y$ samples. An anti-correlation between $T_c$ and $T_{sn}$ is revealed. The data of AFM parent Fe$_{1.19}$Te ($y = 1$) is also shown.

Furthermore, it is clear from Fig. 5 that the superconductivity is significantly suppressed with enhanced $T_{sn}$: The superconducting $T_c$ is reduced from 14 K ($y = 0.61$) to 6.7 K ($y = 0.06$) in FeSe$_{1-y}$Te$_y$, and from 10.8 K ($x = 0.13$) to 4.8 K ($x = 1$) in FeSe$_{1-x}$S$_x$. Based on the related experimental observations and the above discussions, one may take $T_{sn}$ of the induced spin nematicity as a measure of the degree of the anisotropy in AFM correlations to a certain extent. Thus, such a suppression of superconductivity implies that the enhanced anisotropy in AFM fluctuations undermines the superconductivity. This is also consistent with the enhancement of superconductivity in the presence of almost isotropic AFM fluctuations[43,44] as observed in heavily electron-doped FeSe-based intercalate of (Li, Fe)OHFeSe ($T_c$ = 42 K). In other words, the superconductivity is suppressed at least partly due to the reduced contributing pairing channels within the AFM-fluctuation-driven pairing mechanism.

On the other hand, it is worth mentioning that the suppression of superconductivity in these isoelectronic $FeSe_{1-x}S_x$ and $FeSe_{1-y}Te_y$ samples seemingly follows the reduction trend of the electron-electron correlation with increasing $x$ or decreasing $y$ as reported in previous studies.[25,41,42] However, the previous studies have shown that no persistent positive correlation between the superconductivity and electronic correlation exists in the whole regions of $FeSe_{1-x}S_x$ and $FeSe_{1-y}Te_y$ systems, since in general they display a non-monotonic variation of $T_c$ against the continuous reduction of electronic correlation,[25,41,42] though the positive correlation has been shown to hold for heavily electron-doped $Rb_{0.8}Fe_2(Se_{1-z}S_z)_2$ system.[45] The superconductivity of $Rb_{0.8}Fe_2(Se_{1-z}S_z)_2$ is enhanced to a high $T_c$ (32 K) at moderate correlation.[45] Additionally, in prototypical $Fe_{1-x}Se$ and intercalated $(Li, Fe)OHFe_{1-x}Se$ compounds, $T_c$ has been reported as being suppressed with the increased hole doping that is associated with the Fe-deficiency $x$, due to the hole-doping $x$ enhanced electronic correlations.[46,47] Therefore, further investigations are required to clarify the complex effects of electronic correlations on the superconductivity in multi-orbital/band iron-based compounds. Nevertheless, here an anti-correlation between $T_c$ of the superconductivity and $T_{sn}$ of the induced spin nematicity is observed in both $FeSe_{1-x}S_x$ and $FeSe_{1-y}Te_y$ samples (Fig. 5), and the spin correlation has been shown to be very likely separate from the electronic correlation with respect to the isoelectronic substitution in the $FeSe_{1-x}S_x$ system (Fig. 4). Moreover, the induced spin nematicity is found to develop basically from the tetragonal normal-state regime (Fig. 3). Therefore, we conclude that the spin-related physics that originates in the tetragonal regime, rather than the electronic nematic regime, underlies the superconductivity.

In summary, we show the ubiquitous presence of a spin nematicity in isoelectronic $FeSe_{1-x}S_x$ and $FeSe_{1-y}Te_y$ systems, based on systematic measurements of the angular-dependent magnetoresistance in the *ab* plane. This spin nematicity is induced by the in-plane magnetic field $\gtrsim 3$ T and emerges from the tetragonal normal-state regime at a characteristic temperature $T_{sn}$, no matter whether the sample is ordered nematically at $T_s \lesssim T_{sn}$ or magnetically at $T_N < T_{sn}$, or shows neither of the orders. The issue of a possible involvement of the Fe $d_{xy}$ orbital, besides the $d_{xz}/d_{yz}$ orbitals, in the induced spin nematicity is worthy of further investigation. Our results highlight that the isoelectronic substitution with sulfur directly suppresses the spontaneous electronic nematicity, rather than the induced spin nematicity, in the $FeSe_{1-x}S_x$ system. Thus, the spin correlation is very likely separate from the electronic correlation to the certain extent. Furthermore, the enhancement of the induced spin nematicity leads to the significant suppression of the superconductivity. This implies that the enhanced anisotropy in AFM fluctuations undermines the superconductivity, consistent with the AFM-fluctuation-driven pairing scenario. These results provide the new evidence that the spin-related physics originating from the tetragonal background is fundamental to the iron-based unconventional superconductivity.

We thank Profs Jiangping Hu and Kun Jiang for helpful discussions. This work is supported by the National Key Research and Development Program of China (Grant Nos. 2016YFA0300300 and 2017YFA0303003), the National Natural Science Foundation of China (Grant Nos. 12061131005, 11834016 and 11888101), the Strategic Priority Research Program of Chinese Academy of Sciences (Grant Nos. XDB25000000), and the Strategic Priority Research Program and Key Research Program of Frontier Sciences of the Chinese Academy of Sciences (Grant Nos. QYZDY-SSW-SLH001).

*Supplemental Material for*

# Magnetic-Field-Induced Spin Nematicity in FeSe$_{1-x}$S$_x$ and FeSe$_{1-y}$Te$_y$ Superconductor Systems

Shaobo Liu, Jie Yuan, Sheng Ma, Zouyouwei Lu, Yuhang Zhang, Mingwei Ma, Hua Zhang, Kui Jin, Li Yu, Fang Zhou[*], Xiaoli Dong[*], and Zhongxian Zhao

* Correspondence to: fzhou@iphy.ac.cn (F.Z.); dong@iphy.ac.cn (X.D.)

**This file includes the following supplemental figures:**

**Figure S1**: Property characterizations of FeS single crystal grown hydrothermally

**Figure S2**: Temperature-dependent in-plane resistivity and its first derivative for parent Fe$_{1.19}$Te

**Figure S3**: The x-ray $\phi$ scans for FeSe$_{1-x}$S$_x$ ($x = 0.07$) and FeSe$_{1-y}$Te$_y$ ($y = 0.06, 1$)

**Figure S4**: Towfold-anisotropic $\Delta R(\theta,T)/R_{min}(T)$ measured on the same FeSe sample with three different current directions ($\alpha = 0°, 22°, 90°$)

**Figure S5**: Field dependences of the maximum $\Delta R(T)/R_{min}(T)$ at given $T < T_{sn}$ for nematic FeSe$_{0.87}$S$_{0.13}$

**Figure S1**

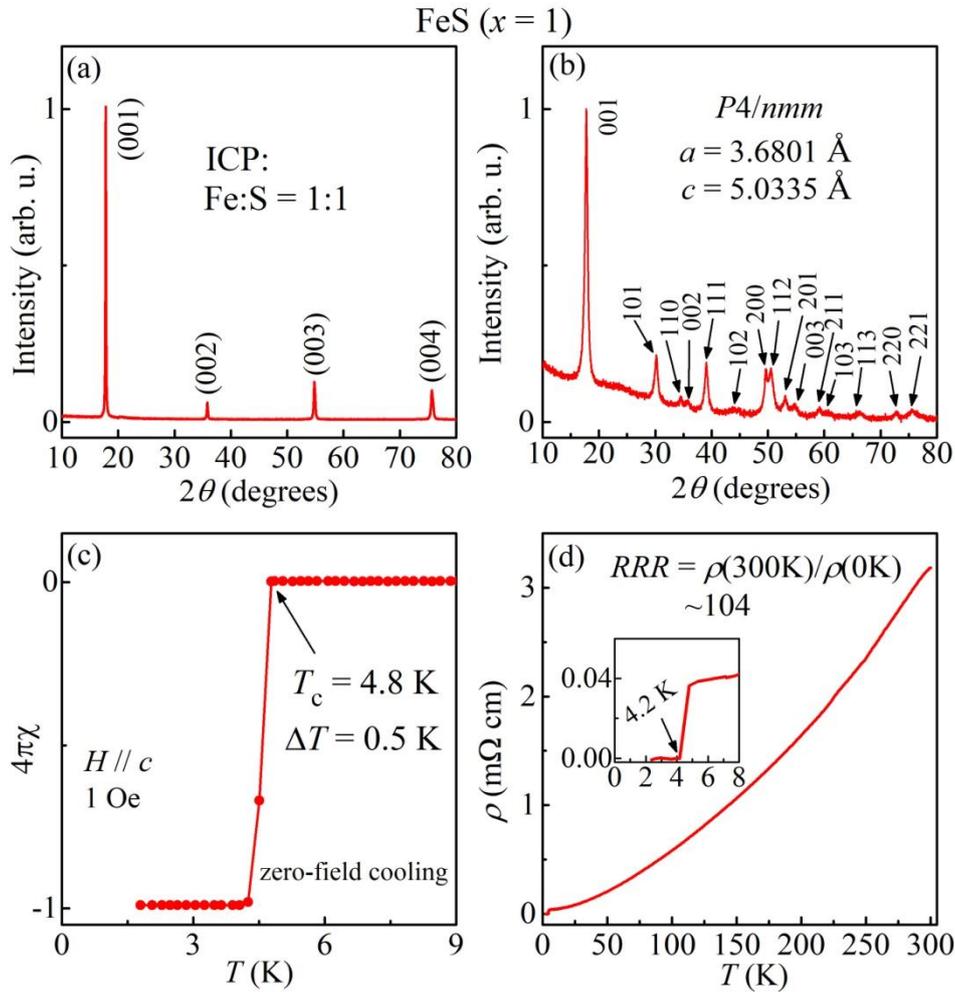

Fig. S1. (a) The single-crystal XRD pattern of the FeS sample demonstrates its single preferred (001) crystal orientation. Its chemical stoichiometry is determined by ICP analysis. Both the single-crystal and powder (b) XRD patterns at room temperature confirm the tetragonal crystal symmetry. The calculated lattice constants, $a$ and $c$, of FeS are the smallest among the FeSe$_{1-x}$S$_x$ and FeSe$_{1-y}$Te$_y$ systems[25], consistent with previous reports. The superconductivity is characterized by both diamagnetism (c) and resistivity (d) measurements.

**Figure S2**

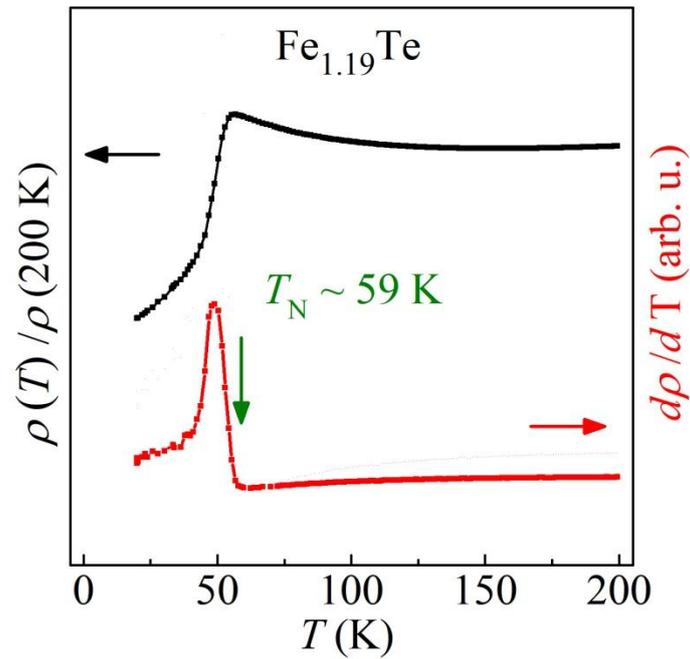

Fig. S2. The temperature-dependent scaled in-plane resistivity and its first derivative of parent $Fe_{1.19}Te$ show the pronounced anomaly at $T_N \sim 59$ K.

**Figure S3**

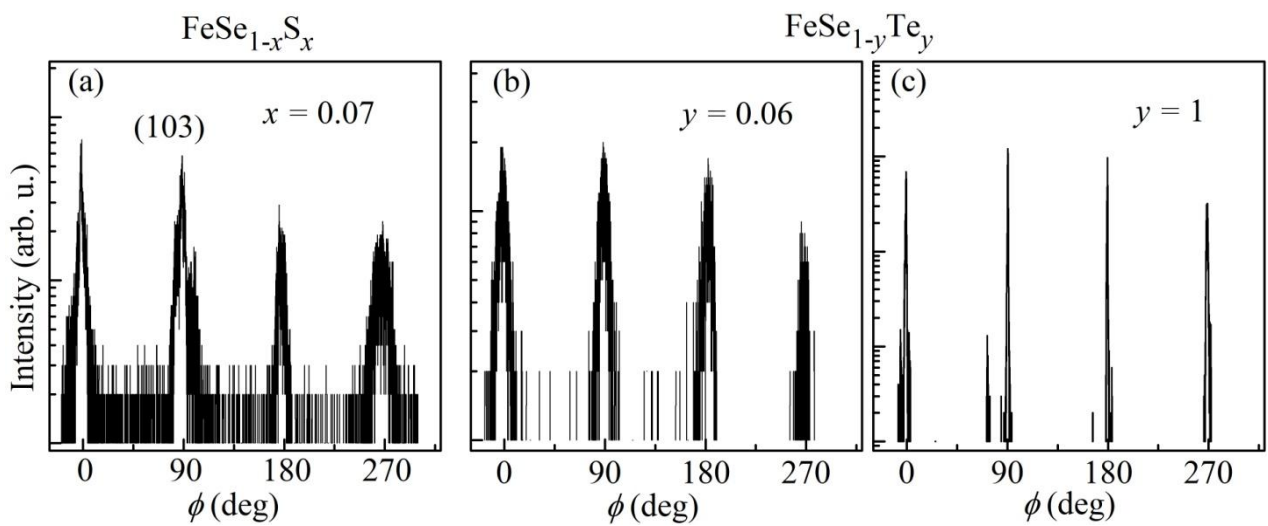

Fig. S3. The room-temperature x-ray $\phi$ scans of (103) plane for (a) $x = 0.07$ $FeSe_{1-x}S_x$, (b) $y = 0.06$ and (c) $y = 1$ $FeSe_{1-y}Te_y$ samples.

**Figure S4**

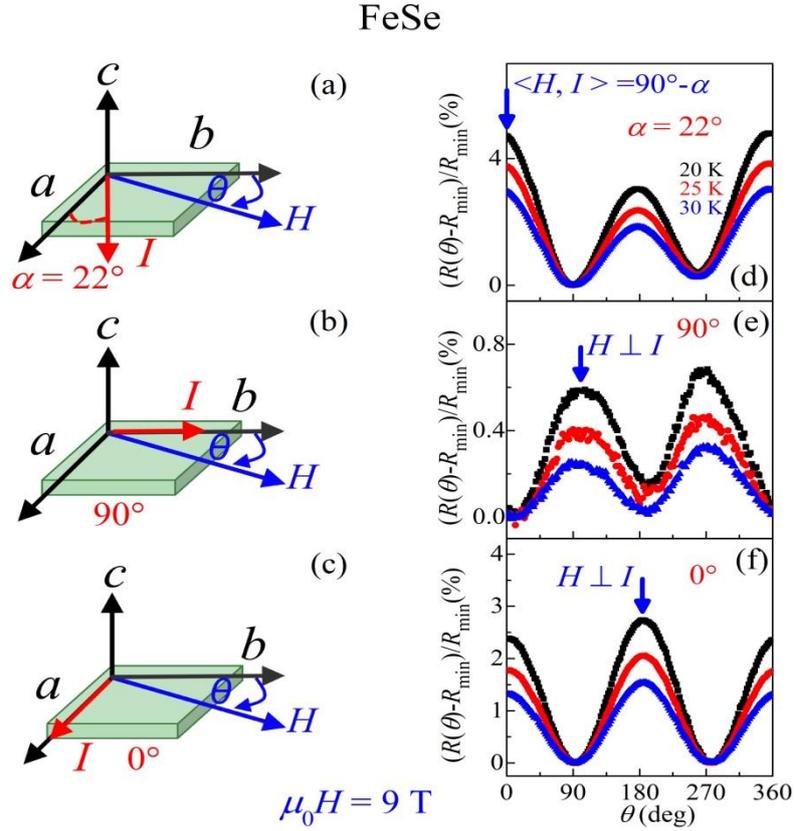

Fig. S4. (a-c) Illustrations of three different current directions ($\alpha$'s) used in the AMR measurements on the same FeSe sample with $T_{sn} \sim 90$ K. (d-f) The corresponding twofold-symmetric $\Delta R(\theta,T)/R_{min}(T)$ of the sample measured under magnetic field of 9 T at $T = 20$ K, 25 K and 30 K.

**Figure S5**

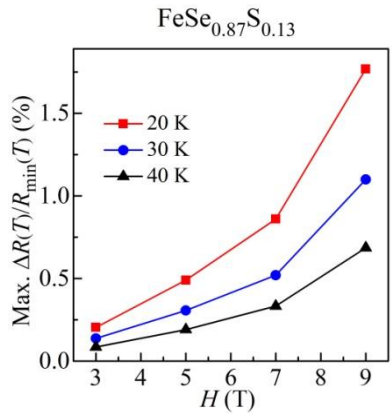

Fig. S5. Field dependences of the maximum $\Delta R(T)/R_{min}(T)$ at $T = 20$ K, 30 K and 40 K for nematic FeSe$_{0.87}$S$_{0.13}$ ($x = 0.13$) with $T_{sn} \sim 90$ K. The data are extracted from Fig. 3(h) in the main text.